\documentclass[manuscript]{aastex}          

\shorttitle{Near-IR observations of GQ Lup b with NIFS}
\shortauthors{Lavigne et al.}

\begin{document}

\title{Near Infrared Observations of GQ Lup b Using the Gemini Integral Field Spectrograph NIFS}

\author{Jean-Fran\c{c}ois Lavigne\altaffilmark{1,2,3} and Ren\'e Doyon}
\affil{D\'epartement de physique and Observatoire du Mont M\'egantic, Universit\'e de Montr\'eal, C.P. 6128, succ. Centre-Ville, Montr\'eal, QC, Canada}
\email{lavigne@astro.umontreal.ca}

\author{David Lafreni\`ere}
\affil{Department of Astronomy and Astrophysics, University of Toronto, 50 St. George Street, Toronto, ON, M5S 3H4, Canada}

\author{Christian Marois}
\affil{NRC Herzberg Institute of Astrophysics, 5071 West Saanich Rd, Victoria, BC, V9E 2E7, Canada}

\and

\author{Travis Barman}
\affil{Lowell Observatory, 1400 West Mars Hill Road, Flagstaff, AZ 86001, USA}

\altaffiltext{1}{Visiting worker at the Herzberg Institute of Astrophysics}
\altaffiltext{2}{Student fellow at the Institut national d'optique}
\altaffiltext{3}{Present address: Institut national d'optique, Parc technologique du Qu\'ebec m\'etropolitain, 2740 rue Einstein, Qu\'ebec, QC, Canada}

\begin{abstract}
We present new $JHK$ spectroscopy ($R \sim 5000$) of GQ Lup b, acquired with the near-infrared integral field spectrograph NIFS and the adaptive optics system ALTAIR at the Gemini North telescope. Angular differential imaging was used in the $J$ and $H$ bands to suppress the speckle noise from GQ Lup A; we show that this approach can provide improvements in signal-to-noise ratio (S/N) by a factor of $2-6$ for companions located at subarcsecond separations. Based on high-quality observations and GAIA synthetic spectra, we estimate the companion effective temperature to $T_{\rm{eff}}$ = $2400 \pm 100$ K, its gravity to $\log g$ = $4.0 \pm 0.5$, and its luminosity to $\log (L/L_\odot)$ = $-2.47 \pm 0.28$. Comparisons with the predictions of the DUSTY evolutionary tracks allow us to constrain the mass of GQ Lup b to $8-60 ~M_{\rm{Jup}}$, most likely in the brown dwarf regime. Compared with the spectra published by Seifahrt and collaborators, our spectra of GQ Lup b are significantly redder (by $15-50$\%) and do not show important $Pa\beta$ emission. Our spectra are in excellent agreement with the lower S/N spectra previously published by McElwain and collaborators. 
\end{abstract}


\section{Introduction}

An important goal of exoplanet imaging is understanding how planetary systems form and this requires observing planets when they are very young, i.e. at ages less than a few million years. Ten years after the first discovery of an extra-solar giant planet (EGP), the imaging era of exoplanet science has finally started. The year 2008 has been particularly prolific in that matter with the discovery of the first planetary system around the $\sim 60$ Myr old HR 8799 \citep{Marois08}, of the exoplanet candidates in orbit around the $100-300$ Myr old Fomalhaut \citep{Kalas08}, in orbit around the $\sim$ 12 Myr old $\beta$ Pictoris \citep{Lagrange08}, the potential companion to the $\sim 5$ Myr old 1RXS J160929-210524 \citep{Lafreniere08} and the planetary mass object (PMO) candidate evolving round the 0.9 to 3 Myr old T-Tauri star CT CHA \citep{Schmidt08}. These new discoveries add to a few previous detections of planet candidates such as the PMO in orbit around the young ($\sim$ 10 Myrs) 25 $M_{\rm{Jup}}$ brown dwarf (BD) 2M1207 \citep{Chauvin04} and the companion in orbit around the $\sim$ $1-10$ Myr old T Tauri star GQ Lup \citep{Neuhauser05}.

GQ Lup b is particularly interesting; the co-moving companion is relatively bright (K=13) and lies at 0.7 arcsec (100 $\pm$ 50 AU) from the 7th magnitude K7eV primary. Despite several studies, the mass of GQ Lup b is still poorly constrained. \citet{Neuhauser05} identified prominent CO and $\rm H_2O$ bands typical of cool M9$-$L4 dwarfs but could not constrain accurately the mass of the companion ($1-42 ~ M_{\rm{Jup}}$) because the effective temperature ($T_{\rm{eff}}$; $1600-2500$ K), the gravity and the distance of the object were not well determined. \citet{Marois07} have analyzed archived HST WFPC2 and NICMOS data and Subaru $\rm CH_4$, $H$, $K_s$ and L$^\prime$ band images of GQ Lup b and provided estimates of its bolometric luminosity ($\log (L/L_\odot)=-2.42 \pm 0.07$ at 140 pc), its radius (0.38 $\pm$ 0.05 $R_\odot$) and its effective temperature (2335 $\pm$ 100 K). Assuming a $\log~ g~ = 3$ and based on the evolutionary models of \citet{Baraffe03}, they inferred a mass ranging from 10 to 20 $M_{\rm{Jup}}$. \citet{McElwain07} secured $J$- and $H$-band spectra and found the companion to be somewhat hotter than previously reported with a spectral type between M6 and L0 corresponding to effective temperatures between 2400 and 2900 K assuming the effective temperature - spectral type relationship from \citet{Golimowski04}. Using the evolutionary models from \citet{Burrows97} and \citet{Chabrier00} and their derived bolometric luminosity of $\log (L/L_\odot)=-2.46 \pm 0.14$ they inferred a mass of $10-40 ~ M_{\rm{Jup}}$. A mass estimate has also been reported by \citet{Seifahrt07} who acquired $J$-, $H$- and $K$-band spectra and compared them to the predictions of the GAIA model from \citet{Brott05}. The best fit resulted in $T_{\rm{eff}} = 2550 - 2750 K$ and $\log~ g~ = ~3.7 \pm 0.5$ dex. From these values, their estimated $\log (L/L_\odot) = -2.25 \pm 0.24$ and by comparison with 2M0535 \citep{Stassun06} which has a dynamically determined mass of $36 \pm 3$ $M_{\rm{Jup}}$ and a similar age as GQ Lup, they derived a mass between 4 and 36 $M_{\rm{Jup}}$. A summary of previously reported physical parameters of GQ Lup b is given in Table \ref{tab:GQLupCarac}.

In light of the International Astronomical Union (IAU) working definition which states that a planet must have a mass below the deuterium burning limit of $\sim 13$ $M_{\rm{Jup}}$, the nature of GQ Lup b thus remains uncertain. Further progress will be made through a better distance estimation of the system and better spectroscopic characterization of the companion.

In this paper, we present new $R \sim 5000$ $JHK$ spectroscopic observations of the GQ Lup system taken with the integral field spectrograph NIFS in combination with the adaptive optics system ALTAIR at the Gemini North telescope using angular differential imaging (ADI) to improve the signal-to-noise ratio (S/N) of the companion. The resulting spectrum is analyzed and compared with synthetic atmospheric spectra and evolutionary tracks to infer the physical parameters of the companion.

\section{Observations}

The GQ Lup system was observed in the $J$, $H$ and $K$ bands respectively on the nights of 2007 May 30, June 1 and June 27 at the Gemini North Telescope (program GN-2007A-Q-46) using NIFS combined with the ALTAIR adaptive optics system \citep{Herriot98}. NIFS is a near-infrared integral field spectrograph (IFS) based on the image slicer concept \citep{McGregor02}. It has a spectral resolving power of 5000 in each band, a total field of view of $3.0\arcsec \times 3.0\arcsec$ and rectangular spatial resolution elements of $0.04\arcsec \times 0.1\arcsec$. 

All observations were obtained with the Cassegrain rotator turned off to allow rotation of the field around the host star. The $J$- and $H$-band sequences consisted of $18 \times 400$s and $27 \times 300$s exposures, respectively, without any dithering. This allowed enough field rotation to use angular differential imaging (ADI; \citet{Marois06,Lafreniere07}) to suppress the speckle noise. As the PSF is better sampled and AO achieves better performance in the $K$ band, observations were obtained with a 9-position square dithering pattern with each position offsetted by $0.4\arcsec$; in total 72 exposures of 50s were obtained. The median seeing was measured to be $0.44\arcsec$, $0.61\arcsec$ and $0.67\arcsec$ in $J$, $H$ and $K$ respectively and the observations were conducted under a median air mass of 1.79 in both $J$ and $H$ and of 1.83 in $K$. A spectroscopic standard for telluric correction was observed at five random dither positions having a maximal excursion of $\pm 0.7\arcsec$ after the $J$-band sequence and both before and after the $H$ and $K$ bands observations. The spectroscopic standards for telluric correction were HIP 82714 in $J$ and HIP 69021 in both $H$ and $K$. Sky and arc lamp frames were taken at the end of each sequence. Darks and flats as well as Ronchi flats allowing the calibration of the spatial distortion of the camera were taken during daytime calibration. No astrometric calibration data (plate scale and orientation) were obtained. An observation log is shown in Table \ref{tab:ObsLog}.

\section{Data Reduction}

\subsection{Primary data reduction and data cube reconstruction}
The data cube reconstruction was done with the IRAF Gemini data reduction package version 1.9.1a. The calibration data were first processed to construct a flat field, a bad pixel map, a wavelength solution and a map of the curved spectral traces. 

The telluric data cubes were then reconstructed. The images were rectified to correct for the trace curvature and re-sampled to yield a common constant wavelength step for all spatial elements. A sky cube was built from the median of the dithered spectroscopic standard data and subtracted from each telluric cube. The latter were then corrected for atmospheric dispersion by fitting a Gaussian kernel to the PSF to find its center in each spectral slice and by translating the PSFs to the center of the field-of-view. The median telluric spectrum was computed and divided by the spectrum of a blackbody at 10 000 K, the approximate A0V spectroscopic standard temperature. 

The final step consisted in reducing the science data and building their associated data cubes. The sky frame obtained after the science observations was first subtracted from each image of the time-lapsed sequence. The science data cubes were extracted following the same procedure used for the telluric data. Telluric correction was then applied. Atmospheric dispersion was then corrected as before. Interstellar extinction was considered to be negligible as most of the extinction associated with GQ Lup A is believed to be circumstellar \citep{McElwain07}. The rectangular IFS spatial resolution elements were interpolated to yield a spaxel size of $0.043\arcsec \times 0.043\arcsec$ in the final reconstructed cube. The term spaxel will be used to refer to a single $(x,y,\lambda)$ element in the final data cube. This is applied to all IFS images to get a final dataset composed of time-lapsed data cubes. 

\begin{figure}[!h]
  \begin{center}
    \includegraphics[width=12cm,height=12cm]{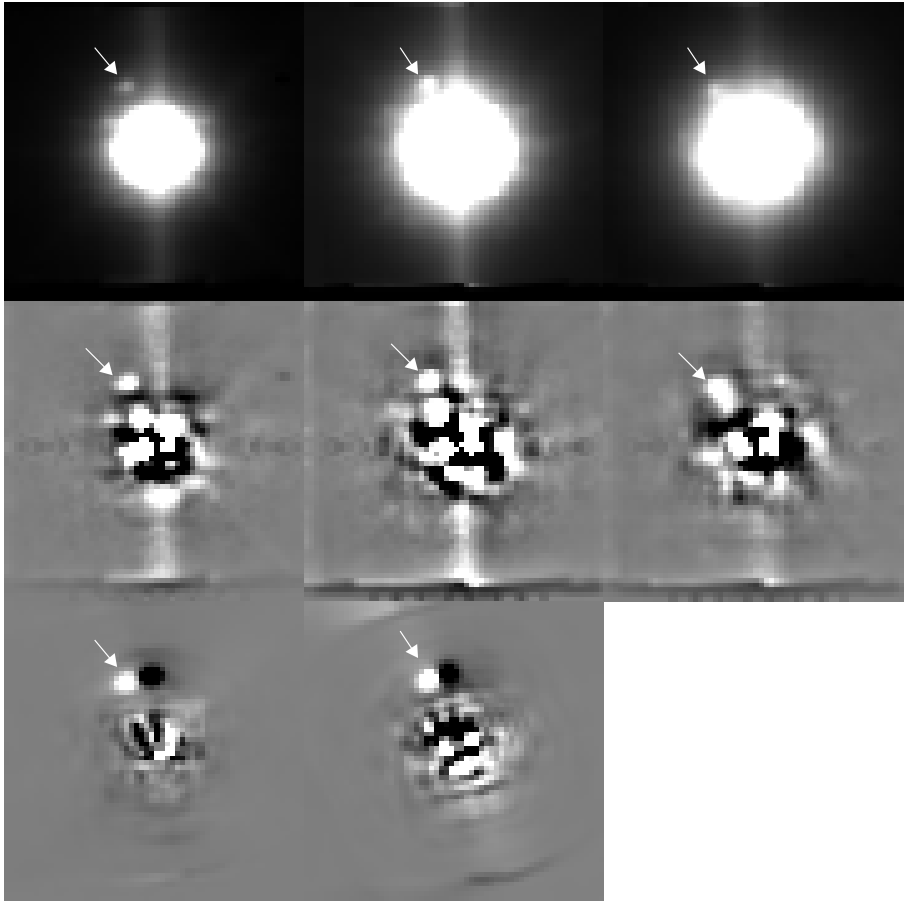}
    \caption{Detection of the companion to GQ Lup in $J$, $H$ and $K$ from left to right. Each image has a field of view of $3.0\arcsec \times 3.0\arcsec$. The top, center and bottom panels respectively show examples of collapsed data cubes before and after the subtraction of a radial profile and after the application of the ADI algorithm. Arrows indicate the companion position. Each image was divided by the host star flux integrated in a two pixels diameter circular aperture and the same dynamic range is used to display all bands. Images after the subtraction of the radial profile and after the application of the ADI algorithm are also shown on the same dynamic range.
    \label{fig:figCompAllBands}}
  \end{center}
\end{figure}

An example of a collapsed data cube, i.e. integrated over the full wavelength range of the photometric band, obtained for each observed band is shown in the top panels of Figure \ref{fig:figCompAllBands}. A bright vertical stripe aligned with NIFS slitlets is present in all our data. NIFS is known to have some scattered light in each independent channel which can result in an enhanced sky proportional to the star brightness in each slitlet and could explain such a bright stripe. 

In preparation for the extraction of the signal of GQ Lup b, the bright signal from the primary PSF was subtracted to first order by applying an unsharp mask with a Gaussian kernel of six pixels FWHM, followed by the subtraction of a median radial intensity profile about the center of the PSF. The images were then convolved by a 1 pixel FWHM Gaussian kernel to attenuate the pixel-to-pixel noise resulting from the interpolation of the rectangular IFS spatial resolution elements to the square data cube spaxels. The resulting residual broad band images obtained after collapsing the data cubes are shown in the lower panels of Figure \ref{fig:figCompAllBands}. Note that the residual vertical stripe is subtracted by the ADI speckle suppression algorithm (see next section) and does not impact the final $J$ and $H$ spectra. In the $K$ band, however, no such algorithm is used. We therefore only considered the first 45 data cubes obtained, in which the companion does not overlap with the vertical stripe. The $K$ band spectrum was extracted from these residual images by summing the flux in a 2 spaxel diameter circular aperture.

\subsection{Angular Differential Imaging}
\label{sec:ADI}
As can be seen in Figure \ref{fig:figCompAllBands}, the residual speckle noise is significant at the separation of the companion and this severely limits the accuracy by which the spectrum can be extracted. In principle, this speckle noise could be attenuated using spectral deconvolution techniques \citep{Sparks02,McElwain07,Thatte07,Janson08,Lavigne09} but, in practice, the companion radial displacement is only $2-3$ NIFS spaxels in a data cube rescaled such that the speckles at different wavelengths overlap. This, combined with the companion brightness, makes an accurate fit of the speckle pattern contaminating the companion spectrum very hard to obtain.

The approach chosen instead was to use the ADI technique developed by \citet{Marois06}. This technique consists in acquiring a series of images with the Cassegrain rotator of an alt-az telescope turned off. This allows field rotation between each observation of the sequence. The companion is then rotating around its host star in the time-lapse sequence while the quasi-static speckles introduced by the defects of the telescope optical elements stay fixed. For each data cube of the time series, a reference data cube can be built from the other data cubes, and used to subtract the speckle pattern and attenuate the speckle noise. 

\begin{figure}[!h]
  \begin{center}
    \includegraphics[width=12cm,height=4cm]{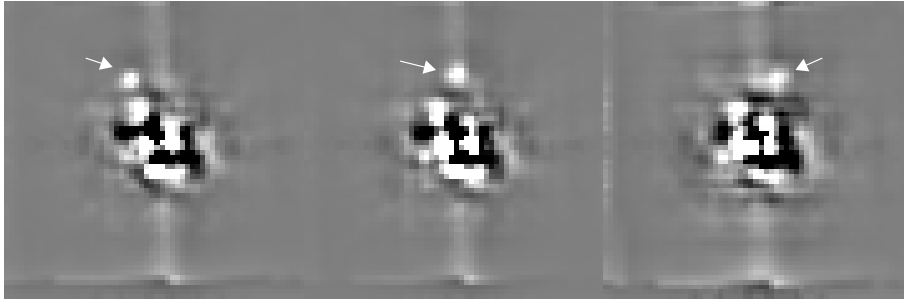}
    \caption{$H$-band time sequence of images with the Cass-rotator turned off. Images are separated by 65 minutes each. The speckles long life time is clearly seen from this sequence. Arrows indicate the companion position. The field has rotated by $21.0^\circ$ in the second image and by $37.7^\circ$ in the third.
    \label{fig:figAdiRot}}
  \end{center}
\end{figure}

Examples of images obtained after the subtraction of their median radial profile taken at intervals of 65 minutes in the $H$ band are shown in Figure \ref{fig:figAdiRot}. This corresponds to an image at the beginning, the middle and the end of the whole sequence. The long speckle life time is clearly seen in that sequence while the companion revolves around its host star due to field rotation.    

The reference data cube was built using the ``locally optimized combination of images'' (LOCI) algorithm developed by \citet{Lafreniere07}. This algorithm builds a reference PSF having the highest speckle correlation with respect to the analyzed frame from the dataset composed of the time sequence of images. This is done by first dividing the frame in optimization subsections. The reference PSF is then built for each zone by computing the linear combination of images that guarantees the lowest possible residual noise after subtraction. Only the images of the time sequence in which the companion had a sufficient displacement compared to the analyzed frame are used to build the reference PSF to avoid a significant companion flux loss. The resulting minimal time separation between two frames is the one yielding the best companion S/N when two frames are differentiated. This was found to be 2400s in $J$ and 2700s in the $H$ band which respectively corresponds to 6 and 9 images in the time sequence and to a minimal companion displacement of 2.7 and 2.9 pixels.

Since the companion position is already known, a single horseshoe shaped optimization zone centered on the host star and having a mean radius corresponding to that of the companion was used. The radial width is taken to be 4 pixels. The missing azimuthal zone is the one in which the companion is moving through the time sequence. This algorithm is applied to each spectral slice of each data cube. The resulting data cubes are then rotated to align the field of view and summed to build the final ADI data cube. The companion spectrum was extracted from this final cube by summing the flux in a 2 spaxel diameter circular aperture.

Applying this ADI process to each data cube has the advantage of increasing the S/N of the companion spectrum by significantly attenuating the speckle noise. However, a fraction of the companion signal is lost through the application of the LOCI algorithm \citep{Lafreniere07}. Moreover, the fact that each spectral slice is treated independently from the others can have an impact on the resulting spectral shape. Hence, we have developed an algorithm to determine the influence of ADI on the companion spectrum.

\begin{figure}[!h]
  \begin{center}
    \includegraphics[width=12cm,height=12cm]{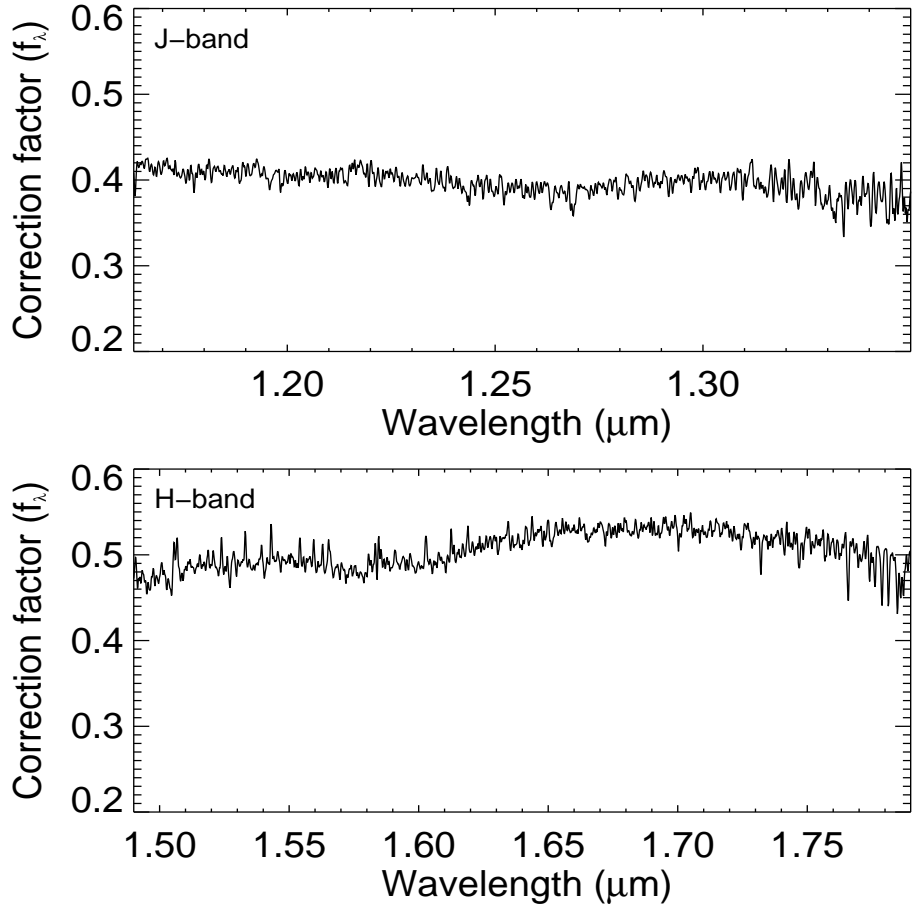}
    \caption{Correction factors applied to the extracted $J$- and $H$-band spectra at a spectral resolution of $R \approx 5000$ to compensate for the flux loss during spatial filtering and the ADI process. The fake companions used to compute this factor had a flat spectrum (see section \ref{sec:ADI} for more details).
    \label{fig:graphSpecCorrection}}
  \end{center}
\end{figure}

For simplicity, let us consider only a single slice of a given data cube in the sequence, keeping in mind that the procedure described below is repeated for each slice and each cube. Following the execution of the above LOCI subtraction algorithm, the residual integrated flux at the companion position in the cube considered can be written as
\begin{equation}
F=I-\sum_i c_i I_i,
\end{equation}
where $I$ is the integrated flux in the slice considered before LOCI subtraction, $I_i$ is the same but for cube $i$, and $\{c_i\}$ is the set of coefficients found by the LOCI algorithm. The integrated flux $I$ can be expressed as the sum of a signal $C$ coming from the companion and a signal $S$ coming from quasi-static speckles, yielding
\begin{equation}
F=\left(S+C\right)-\sum_i c_i \left(S_i+C_i\right)=S_{\rm res} + C-\sum_i c_i C_i,
\end{equation}
where $S_{\rm res}=S-\sum_i c_i S_i$ is the residual speckle signal, which on average will be close to zero. Our goal is to find the true companion signal $C$, and so we must estimate the quantity $\sum_i c_i C_i$, i.e. the fraction of the companion signal that was subtracted by the algorithm. This was done by introducing a fake companion of known flux and flat spectrum into the images, at the same separation as GQ Lup b but 180\degr\ from it. The fake companion was taken to be a shifted and intensity scaled-down version of the primary star; its PSF should thus be very similar to that of GQ Lup b. The LOCI algorithm was applied to the sequence of data with the fake companion by using the exact same coefficients as for the data without the fake companion. Then the fake companion residual flux $F^\prime$ was retrieved. Since the true fake companion signal $C^\prime$ and the residual speckle signal $S^\prime_{\rm res}$ at the location of the fake companion are known (the latter from the execution of the algorithm without the fake companion), the fraction of the fake companion flux that was subtracted by the algorithm in a given spectral slice can be readily computed as
\begin{equation}
f_{\lambda} \equiv \frac{\sum_i c_i C^\prime_i}{C^\prime} = \frac{S^\prime_{\rm res}+C^\prime-F^\prime}{C^\prime}.
\end{equation}
This fraction should be the same for GQ Lup b as for the fake companion since the same coefficients were used for both and they have the same PSF. Note also that $f_{\lambda}$ is independent of the fake companion azimuthal position as $\sum_i c_i C^\prime_i$ only depends on the relative angular position of the companion PSF between the analyzed frame and the cube $i$. So, the true signal of GQ Lup b can be calculated as
\begin{equation}
C = \frac{F-S_{\rm res}}{1-f_{\lambda}},
\end{equation}
where in practice $S_{\rm res}$ is neglected; the resulting error on $C$ is computed by measuring the residual speckle noise at the companion radius. The correction factors $f_{\lambda}$ computed for the $J$ and $H$ ADI spectra by the above procedure are shown in Figure~\ref{fig:graphSpecCorrection}.

Finally, to make sure that the above procedure did not introduce any artificial feature in the spectra, mainly in their continuum shape, we have performed a simple spectrum extraction directly from the reduced data cubes without any ADI processing. The resulting spectra were compared with the spectra obtained with ADI processing and no significant difference was found, save for a lower S/N. More precisely, a similar procedure as for the $K$ band data was used to extract a non-ADI spectrum from the data cubes in which the companion is the least affected by the vertical stripe in the $J$ and $H$ bands. The resulting ADI spectrum was then divided by the non-ADI one to characterize the continuum dispersion. The median non-ADI $J$ band spectrum of the first six data cubes had only a 5\% continuum difference with the ADI spectrum. Similarly, the first eight and last four data cubes in the $H$-band were used to compute the median non-ADI spectrum and the resulting continuum had only a 6\% slope difference with the ADI spectrum.

\section{Results}

\subsection{Speckle noise attenuation performance}
The ability of the algorithms described in the previous section to attenuate the speckle noise in NIFS data is evaluated in this section. The speckle noise attenuation factor is defined as the ratio of the noise in the initial image over the noise in an ADI image. The noise in ADI images was corrected for the companion flux loss during the ADI process using fake companion implants. Each spectral slice was evaluated independently. We report the median value of speckle noise attenuation in all the spectral slices as a function of angular separation from GQ Lup. The noise was evaluated by computing the standard deviation of spaxels within an annulus of one spaxel width at each radius. GQ Lup b was masked out from the images. 

\begin{figure}[!h]
  \begin{center}
    \includegraphics[width=12cm,height=12cm]{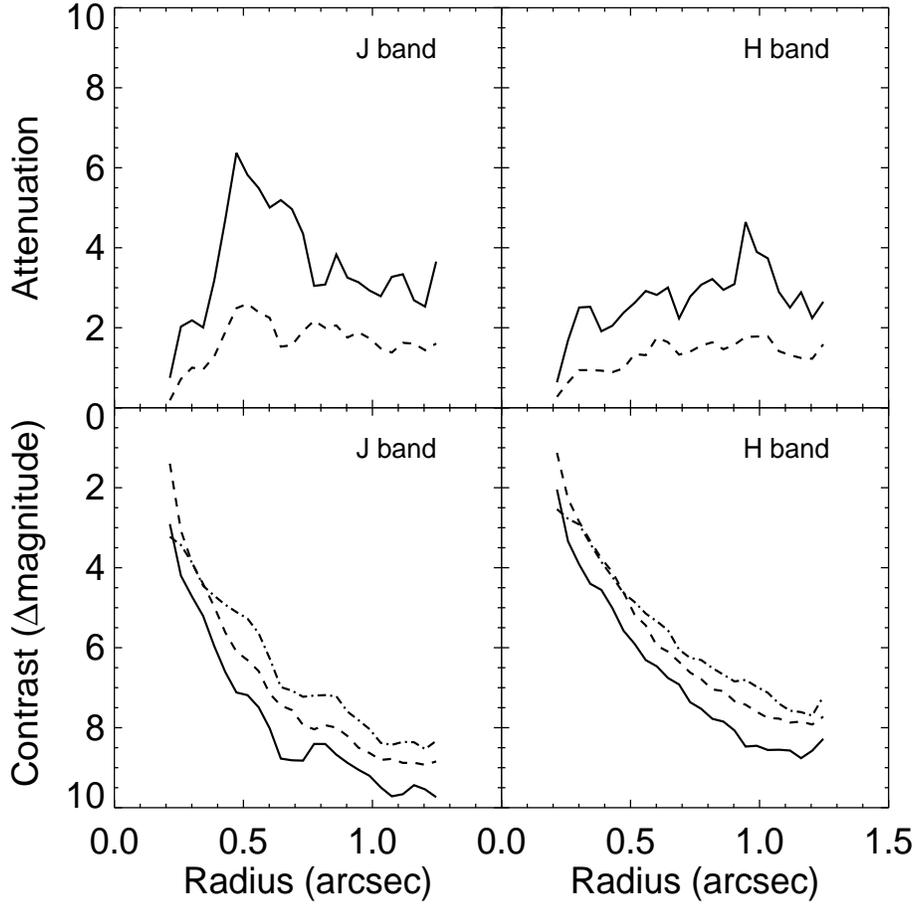}
    \caption{The speckle noise attenuation performance is plotted in the upper panels for collapsed data cubes in $J$ and $H$ bands. The dashed line represents the improvement between a single image of the time sequence and its associated ADI image. The solid line is the attenuation obtained between a single image and the final ADI image constructed by aligning the field of view and adding all the single ADI images. The bottom panels show the median detection limit in the non-ADI images (dot-dashed line), the median detection limit in single ADI images (dashed line) and after field of view alignment and addition of all the single ADI images (solid line). The contrast or detection limit is defined as the $5\sigma$ speckle noise level in an image. The ADI curves were corrected for the companion flux loss resulting from the reference image subtraction. Note that the reference image was constructed with a singular LOCI optimization zone situated at the companion radius and, hence, the reported curves are the lowest achievable performance with NIFS at other radii.
    \label{fig:SpecNoiseAttADI}}
  \end{center}
\end{figure}

The results are plotted at the top of Figure \ref{fig:SpecNoiseAttADI} for the $J$ and $H$ bands while the contrast curves achieved are plotted at the bottom of the figure. For a single data cube, the ADI subtraction provided a noise attenuation by a factor of 1.5 to 2 in both $J$ and $H$, while the combination of all ADI data cubes after alignment of their field of view provided an additional noise attenuation by a factor of 2 to 3. Overall, the speckle noise is attenuated by a factor of $3-6$ in $J$ and a factor $2-5$ in $H$. 

\subsection{Photometry}
The relative photometry of the companion compared with its host star is first evaluated. Unfortunately, the peak of GQ Lup A is saturated or in the detector non-linear regime over most of the $J$ band and the entire $H$ band, preventing us from calculating differential photometry over those bands. In $J$, over the narrow spectral range in which GQ Lup A is in the detector linear regime ($1.31-1.35 ~ \mu m$), we calculated a contrast of $6.53 \pm 0.06$ mag for the companion.

The uncertainty was measured by computing the standard deviation of the flux in 20 apertures identical to the one used to integrate the companion flux at the same radius as the companion. The peak of GQ Lup A is below non-linearity in the entire $K$ band; we measured a differential photometry of $\Delta K_s = 6.35 \pm 0.11$ mag.
 
GQ Lup A is a known variable classical T Tauri star (e.g. \citet{Batalha01}). \citet{Broeg07} reported it to vary by $\pm 0.44$ mag and $\pm 0.22$ mag in $J$ and $K_s$ bands  respectively over a period of $8.38 \pm 0.2$ days while \citet{Neuhauser08} found it to vary by a maximal amplitude of $\pm 0.20$ mag in the $K_s$ band between May 2005 and Feb 2007. Using the GQ Lup A photometry reported by \citet{Neuhauser08} of $K_s = 7.18 \pm 0.20$ mag and the $J$-band photometry from 2MASS $J = 8.605 \pm 0.44$ with the updated error bar from \citet{Broeg07}, we find a GQ Lup b photometry of $K_s=13.45 \pm 0.25$ and $J = 15.13 \pm 0.44$. The $K_s$ value is consistent with $K_s=13.39 \pm 0.08$ from \citet{Neuhauser08} and the $J$-band value is consistent with $14.90 \pm 0.11$ reported by \citet{McElwain07}.

\subsection{Spectroscopy}
Our spectra of GQ Lup b are shown in Fig. 5. The S/N are 145, 76 and 15 at $R \approx 5000$ in $JHK$ respectively where the noise was computed from the dispersion of the flux measured in 20 apertures situated at the same radius as the companion. The integrated flux in each aperture were corrected by $f_\lambda$ (see section \ref{sec:ADI}). 

In the same figure, our $JHK$ spectrum of GQ Lup b is compared with the previously published results obtained by \citet{Seifahrt07} and \citet{McElwain07}\footnote{Note that the \citet{McElwain07} spectra were taken during the OSIRIS commissioning run. The lower S/N is not representative of the camera performance as several adjustments were made to the camera after these observations were taken.}. The first noted feature is that our spectra are redder than the ones reported by \citet{Seifahrt07} in all three bands while they have a similar continuum slope as the ones reported by \citet{McElwain07}. Dividing our spectra by the ones reported by \citet{Seifahrt07} reveals differences in the overall spectral slopes of 22\%, 48\% and 15\% respectively in $J$, $H$ and $K$. This difference is of 8\% in $J$ and of 17\% in $H$ when compared to \citet{McElwain07} where most of the $H$-band difference lies in the bluer spectral range with a 14\% continuum difference below $1.63$ $\mu m$. The difference between the two spectra at the redder end is then of only $\sim 3\%$.  
 
A second difference is seen for the $Pa\beta$ emission line, which is weaker by about an order of magnitude in our spectrum compared to that of \citet{Seifahrt07}. While it is detected with an equivalent width of EW($Pa\beta)$ = $-3.83 \pm 0.15$ by these authors, it is detected with an EW$(Pa\beta)$ = $-0.46 \pm 0.08$ in ours. Emission-lines of PMS accreting objects are known to vary either due to accretion variability, to projection effects on the accreting column rotating around the object or to hot spots that are not always visible to the observer due to the object rotation (see \citet{Scholz06} and reference therein for more details). The strongest $Pa \beta$ variations found in the literature were reported by \citet{Natta06} who analyzed a sample of 14 BDs also observed by \citet{Gatti06} one to two years later. They found a variation of a factor of two at most for all the objects in their sample except  for one that varied by a factor of three. This makes GQ Lup b significantly more variable than any other BD previously observed.

This fact along with the continuum differences raise the possibility that GQ Lup b spectra from \citet{Seifahrt07} are contaminated by some host star flux. Indeed, such a contamination could explain both the bluer continuum and the significantly stronger $Pa\beta$ emission line measured by \citet{Seifahrt07} since the spectrum of A is bluer than B and exhibits strong $Pa\beta$ emission. However, the similarities between the KI doublets EW reported in Table \ref{tab:EwComp} for both spectra argues against this. 

The use of PSF subtraction algorithms could potentially introduce artefacts in the companion spectrum. The fact that different algorithms were used in the three compared papers could then explain the discrepancy between their spectra. As discussed before, we do not believe that artefacts were introduced by our ADI subtraction procedure given the small difference (respectively, 5 and 6\% for the $J$ and the $H$ band) between the median non-ADI and ADI spectra. This assertion is reinforced by the presence of the same discrepancy between our results and the ones from \citet{Seifahrt07} in the $K$ band where no ADI processing was done. At this point, we do not have any satisfactory explanation for the discrepancy between the results of the different papers.

\section{Discussion}
In Figure \ref{fig:graphSpecComp}, our spectra are compared with the $\sim 5$ Myr late-M to early-L dwarfs USco J160830-233511, USco J161302-212428 and USco J163919-253409 from \citet{Lodieu08}\footnote{The spectra can be found at http:/www.iac.es/galeria/nlodieu/}, the $1-50$ Myr 2MASS J014158-463357 \citep{Kirkpatrick06} and some older field dwarfs taken from the NIRSPEC Brown Dwarf Spectroscopic Survey (BDSS) \citep{McLean03}\footnote{The spectra can be found at http:/www.astro.ucla.edu/$\sim$mclean/BDSSarchive/} and from \citet{Cushing05}\footnote{The spectra can be found at http://irtfweb.ifa.hawaii.edu/$\sim$spex/spexlibrary/IRTFlibrary.html}. A first feature that is seen by comparing GQ Lup b with the field dwarfs is that the absorption from $\rm H_2O$ at 1.34 $\mu m$, known as being mostly temperature dependent in low-mass objects (e.g. \citet{Gorlova03}), is similar to objects of spectral types between L0 and L2. This is also supported by the comparison of the continuum slope shortward of $\sim 1.55$ $\mu m$ and longward of $\sim 1.68~\mu m$, while the region in between has been reported as being particularly affected by the reduced $\rm H_2$ collisionally-induced absorption (CIA) in young, low gravity objects such as GQ Lup b \citep{Borysow97,Kirkpatrick06}. This also explains the better agreement in that zone with the younger objects. $\rm H_2$ CIA has its strongest effect in the $K$ band where GQ Lup b does not exhibit the decrease in flux seen between 2.15 and 2.30 $\mu m$ in field objects and has significantly shallower CO absorption. The $K$-band spectrum is better reproduced by the young objects, with the L1 spectra being the most similar.

Hence, we find a spectral type of L1$\pm 1$ for this object. This is at the cooler end of the M6-L0 spectral type reported by \citet{McElwain07} and is consistent with the M9-L4 spectral type inferred by \citet{Neuhauser05}. This is verified by computing the $\rm H_2O$ index at 1.5 $\mu m$ defined by \citet{Allers07}, which is believed to be independent of surface gravity. We find a spectral type of L0 $\pm 1$ also consistent with our spectral type estimate. The $T_{\rm{eff}}$ - spectral type relationship of \citet{Golimowski04} implies $T_{\rm{eff}}=2250 \pm 134~ K$. However, this relation was only calibrated for evolved field objects and thus should be used with caution with young objects. \citet{Kirkpatrick08} also demonstrate that objects of a similar optically determined spectral type exhibit a significant dispersion in their near-IR relative color. This illustrates the challenge of using such a relation to infer $T_{\rm{eff}}$, especially from near-IR data. 

One can also estimate the effective temperature of GQ Lup b and its gravity by comparing our $JHK$ spectra with synthetic ones generated by the GAIA model v.2.6.1 \citep{Brott05}. The grid explored includes R=5000 spectra ranging from $T_{\rm{eff}}$ = 2000 to 2900 K and from $\log~ g~ =$ 1.5 to 5.5 dex with increments of 100 K and 0.5 dex respectively. A $\chi^2$ minimization algorithm was used to find the best fitting parameters. As reported by \citet{Kirkpatrick06}, current atmosphere models poorly reproduce the strong $\rm H_2O$ absorptions between 1.32 and 1.60 $\mu m$ and between 1.75 and 2.20 $\mu m$. Hence, the $H$ band was not used in the fit and those zones were masked out of the $J$- and $K$-band spectra. The $Pa\beta$ emission line in the $J$ band was also masked out of the fit. The $K$-band fit is poor and leads to an optimal $T_{\rm{eff}}$ $>$ 2900 K which is largely inconsistent with any previous $T_{\rm{eff}}$ estimates for this object. However, our $K$-band spectrum is very similar to other young brown dwarfs of similar spectral types in the zone fitted as seen in Figure \ref{fig:graphSpecComp}. This indicates that there is still some discrepancy between models and young low-gravity objects in that band. Hence, only the $J$-band spectrum was used to estimate $T_{\rm{eff}}$ and $\log~ g$. Good fits were obtained for $T_{\rm{eff}}$ = $2300 - 2500$ K and $\log~ g$ = $3.5 - 4.5$ with a best fit value of $T_{\rm{eff}}$ = 2400 K and $\log~ g = 4.0$. The model and observed spectra are compared in Figure \ref{fig:GaiaFitSpec} for the $J$ band. The observed $H$- and $K$-band spectra as well as the synthetic ones obtained for those best fitting values are shown in Figure \ref{fig:GaiaFitHK}. Hence, we adopt $T_{\rm{eff}}$ = $2400 \pm 100$ K and $\log~ g = 4.0 \pm 0.5$ dex. The same exercise was repeated with grids of AMES-dusty and AMES-cond synthetic spectra and all converged to the same solution. 

From these results, the mass of GQ Lup b is evaluated following the procedure used in \citet{Marois08}. The model spectra flux is first adjusted to best fit the photometric data points reported in this work and in the literature through a $\chi^2$ minimization algorithm. The resulting spectra along with the photometric measurements used for the fit are plotted in Figure \ref{fig:compMag}. The luminosity is then evaluated to be equal to $\log(L/L_\odot)=-2.47 \pm 0.28$ dex by integrating the model flux over the whole wavelength range and by using the companion distance estimate from \citet{Neuhauser08} of $139 \pm 45$ pc. Adopting the \citet{McElwain07} age estimate of $1-10$ Myr for the companion and plotting the new luminosity on the DUSTY evolutionary tracks from \citet{Chabrier00}\footnote{The tracks were retrieved from http://perso.ens-lyon.fr/isabelle.baraffe/} yields a mass estimate between $\sim 11-60 ~ M_{\rm{Jup}}$ (see Figure \ref{fig:dustyLum}) putting GQ Lup b most likely in the brown dwarf regime.

Even though evolutionary models are uncertain at the young age of GQ Lup \citep{Baraffe02}, \citet{Stassun06,Stassun07} provide a first anchor point with the $1^{+2}_{-1}$ Myr eclipsing binary 2M0535-05. The lower mass object of this system ($36 \pm 3$ $M_{\rm{Jup}}$) is particularly interesting as it has a mass comparable to the one estimated for GQ Lup b. The comparison of its dynamically measured mass with the prediction of the evolutionary model gives an estimate of the model systematic error. \citet{Chabrier00} evolutionary tracks predict a luminosity of $\log(L/L_\odot)=-2.00$ for an object with such a mass and then underestimate it by a factor of $\sim 1.5$. Even though this lies only at $\sim 1.5\sigma$ of its measured luminosity and, hence, does not constitute a convincing evidence of a model systematic error, it is still interesting to include this error in our analysis to take into account the model uncertainty. Updating our mass estimate with this systematic error leads to a revised lower mass value of $\sim 8~M_{\rm{Jup}}$ which is below the boundary between a planetary mass object and a brown dwarf. Our final mass estimate is then of $8-60 ~ M_{\rm{Jup}}$ for GQ Lup b.

\section{Conclusions}
In this paper, we first derived a method to characterize and compensate for the effect of the ADI speckle suppression algorithm applied to IFS data. We found that the ADI process provided a speckle noise attenuation by a factor greater than 3 in the $J$ band and greater than 2 in the $H$ band. We extracted spectra in the $J$, $H$ and $K$ bands for the low-mass companion GQ Lup b. The companion $Pa\beta$ emission line was marginally detected. The extracted spectra allowed us to constrain its spectral type to $L1 \pm 1$ and to re-evaluate its $T_{\rm{eff}}$ to $2400 \pm 100$ K and $\log$ g to $3.5 - 4.5$. This leads to the re-evaluation of the bolometric luminosity to $\log(L_{bol}/L_\odot)=-2.47 \pm 0.28$ and to a mass estimate of $8-60 ~ M_{\rm{Jup}}$ by comparison to predictions from evolutionary models. This defines GQ Lup b as being most likely a brown dwarf. 

\acknowledgments 
Based on observations obtained at the Gemini Observatory, which is operated by the Association of Universities for Research in Astronomy, Inc, under a cooperative agreement with the NSF on behalf of the Gemini partnership: the National Science Foundation (United States), the Science and Technology Facilities Council (United Kingdom), the National Research Council (Canada), CONICYT (Chile), the Australian Research Council (Australia), Minist\'erio da Ci\^encia e Tecnologia (Brazil) and SECYT (Argentina).

We would like to thank Andreas Seifahrt, Ralph Neuh\"auser, Michael McElwain, James Larkin and J. Davy Kirkpatrick for sharing their spectra with us. We would also like to acknowledge Eric Steinbring at the Canadian Gemini Office for his support during the time proposal redaction and during our data reduction process.

\newpage

\begin{deluxetable}{lccccc}
  \tabletypesize{\scriptsize}
  \tablecaption{GQ Lup b previously published results  \label{tab:GQLupCarac}}
  \tablewidth{0pt}
  \tablehead{
  \colhead{} & \colhead{\citet{Neuhauser05}} & \colhead{\citet{Marois07}} & \colhead{\citet{McElwain07}} & \colhead{\citet{Seifahrt07}} & \colhead{This work}
}
  \startdata
  Age (Myr): & $1 \pm 1$ & $1 \pm 1$ & $1-10$ & $1 \pm 1$ & $1-10$  \\
  Distance (pc): & $140 \pm 50$ & $140 \pm 50$ & $150 \pm 20$ & $140 \pm 50$ & $139 \pm 45$ \\
  $T_{\rm{eff}}$ (K): & $2050 \pm 450$ & $2335 \pm 100$ & $2600 \pm 300$ & $2650 \pm 100$ & $2400 \pm 100$ \\
  $\log$ g: & $2.52 \pm 0.77$ & $3$ (assumed) & - & $3.7 \pm 0.5$ & $4.0 \pm 0.5$ \\ 
  Spectral type: & M9 - L4 & - & M6 - L0 & - & L1$\pm 1$ \\
  Radius $(R_{\rm{Jup}})$: & $\sim 2$ & $3.7 \pm 0.5$ & - & $3.50_{-1.03}^{+1.50}$ & - \\
  $\log(L/L_{\odot})$: & $-2.37 \pm 0.41$ & $-2.42 \pm 0.07$ & $-2.46 \pm 0.14$ & $-2.25 \pm 0.24$ & $-2.47 \pm 0.28$ \\
  Mass $(M_{\rm{Jup}})$: & $1-42$ & $10-20$ & $10-40$ & $4-36$ & $8-60$\\
  \enddata
\end{deluxetable}

\begin{deluxetable}{l l c c c c c c}
  \tabletypesize{\scriptsize}
  \tablecaption{Observation log  \label{tab:ObsLog}}
  \tablewidth{0pt}
  \tablehead{
  \colhead{Band} & \colhead{Date} & \colhead{Number of} & \colhead{Exposure} & \colhead{Median} & \colhead{Median} & \colhead{Total field} & \colhead{Dither} \\
   & & \colhead{exposures} & \colhead{time (s)} & \colhead{seeing ($\arcsec$)} & \colhead{airmass} & \colhead{rotation ($^\circ$)} & ($\arcsec$)  
}
  \startdata
   $J$ band: & 2007 May 30 & 18 & 400 & $0.44$ & 1.79 & $34.3$ & No  \\
   $H$ band: & 2007 Jun 1 & 27 & 300 & $0.61$ & 1.79 & $38.8$ & No  \\
   $K$ band: & 2007 Jun 27 & 72 & 50 & $0.67$ & 1.83 & $33.1$ & $\pm 0.7$  \\
   \enddata
\end{deluxetable}

\begin{deluxetable}{l c c c c c}
  \tabletypesize{\scriptsize}
  \tablecaption{Emission/absorption line equivalent width comparison  \label{tab:EwComp}}
  \tablewidth{0pt}
  \tablehead{
  \colhead{Source} & \colhead{$EW[Pa\beta]$} & \colhead{EW[KI($1.169$ $\mu m$)]} & \colhead{EW[KI($1.178$ $\mu m$)]} & \colhead{EW[KI($1.244$ $\mu m$)]} & \colhead{EW[KI($1.253$ $\mu m$)]} \\
   & \colhead{$[\AA]$} & \colhead{$[\AA]$} & \colhead{$[\AA]$} & \colhead{$[\AA]$} & \colhead{$[\AA]$}
}

  \startdata
   This work: & $-0.46 \pm 0.08$ & $2.12 \pm 0.32$ & $2.97 \pm 0.44$ & $2.30 \pm 0.26$ & $1.40 \pm 0.17$  \\
   \citet{Seifahrt07}: & $-3.83 \pm 0.12$ & $2.22 \pm 0.19$ & $2.51 \pm 0.40$ & $1.72 \pm 0.15$ & $1.61 \pm 0.17$ \\
   \enddata
\end{deluxetable}

\clearpage

\begin{figure}[!h]
  \begin{center}
    \includegraphics[width=10cm,height=12cm]{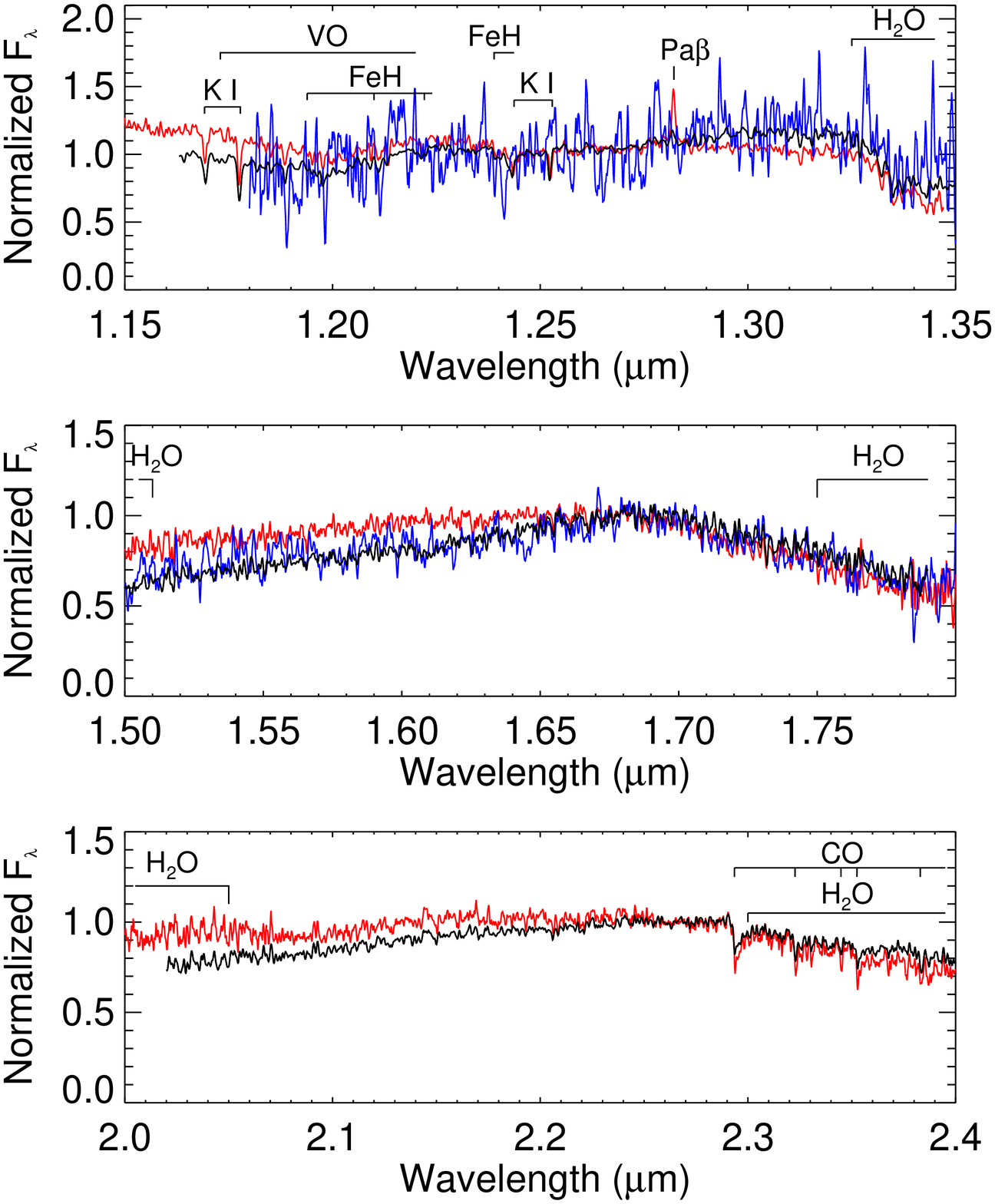}
    \caption{Comparison of GQ Lup b $JHK$ spectra obtained in this work (black lines) with the ones previously published by \citet{Seifahrt07} (red lines) and by \citet{McElwain07} (blue lines). The NIFS spectra were convolved by a Gaussian kernel to get similar spectral resolution as in \citet{Seifahrt07}, namely 2500, 4000 and 4000 in $JHK$ respectively. The \citet{McElwain07} spectra are displayed at their full resolution of $\approx 2000$ in $J$ and $H$. The main absorption/emission features are shown for each band. The spectra are normalized at 1.245, 1.68 and 2.29 $\mu m$ respectively in $JHK$. 
    \label{fig:graphSpecSeifahrt}}
  \end{center}
\end{figure}

\begin{figure}[!h]
  \begin{center}
    \includegraphics[width=12cm,height=12cm]{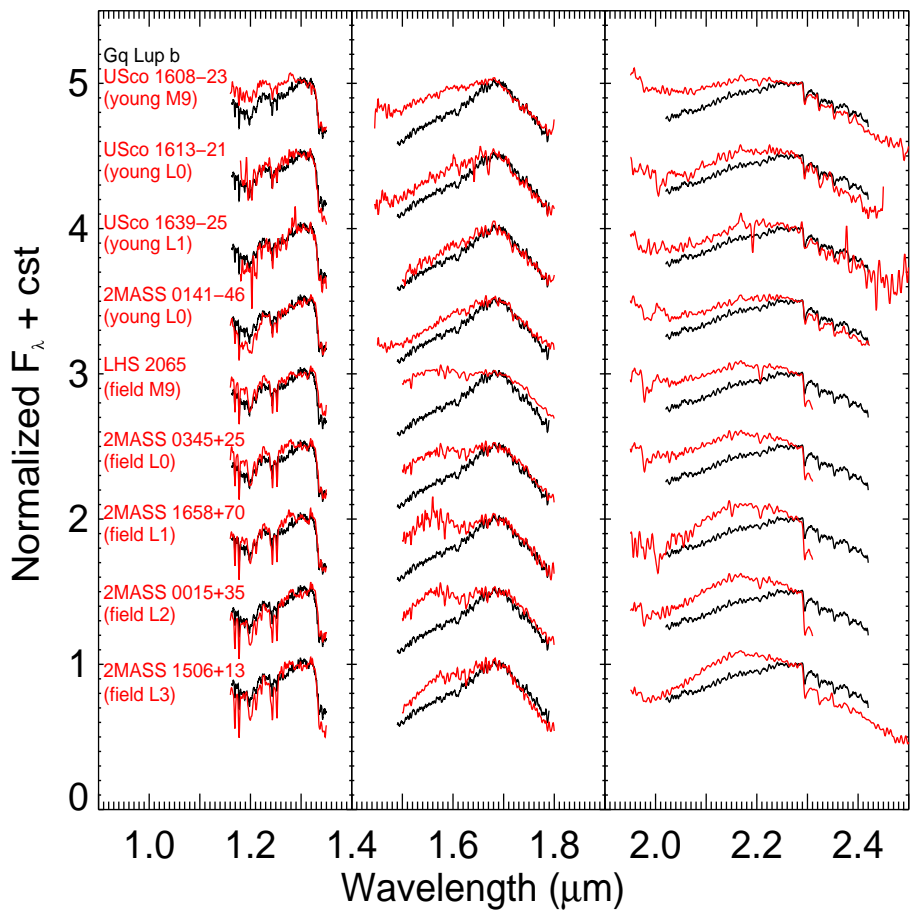}
    \caption{Comparison of GQ Lup b spectra with other known young and field brown dwarfs. The $J$-, $H$- and $K$-band spectra were respectively normalized with respect to their flux at 1.29, 1.68 and 2.29 $\mu m$. The spectra were convolved with a Gaussian kernel to produce spectra at $R \approx 850$. A constant was added to the different spectra for clarity.
    \label{fig:graphSpecComp}}
  \end{center}
\end{figure}

\begin{figure}[!h]
  \begin{center}
    \includegraphics[width=12cm,height=8cm]{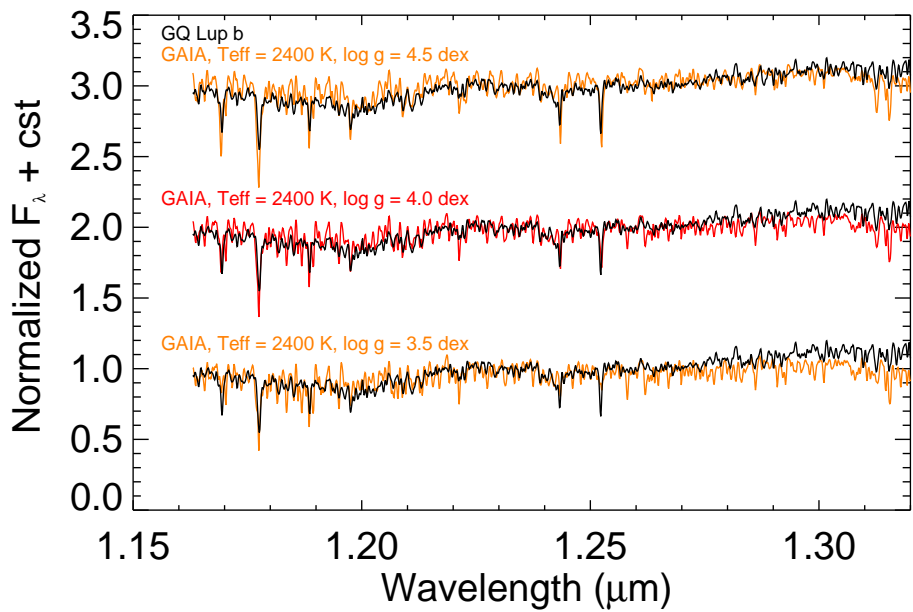}
     \caption{Comparison of the GQ Lup b $J$-band spectra with spectra generated with the GAIA model at $R~\approx ~5000$. Good fits found by a $\chi^2$ minimisation algorithm were obtained for $T_{\rm{eff}}$ = 2300 - 2500 K and $\log$ g = 3.5 - 4.5 dex with a best fit for $T_{\rm{eff}}$ = 2400 K and $\log ~ g$ = 4.0 dex. The model spectra corresponding to the three gravity grid points at 2400 K are displayed as orange lines with the best fit value as the red line. The observed spectrum is depicted as the black lines.  
    \label{fig:GaiaFitSpec}}
  \end{center}
\end{figure}

\begin{figure}[!h]
  \begin{center}
    \includegraphics[width=10cm,height=12cm]{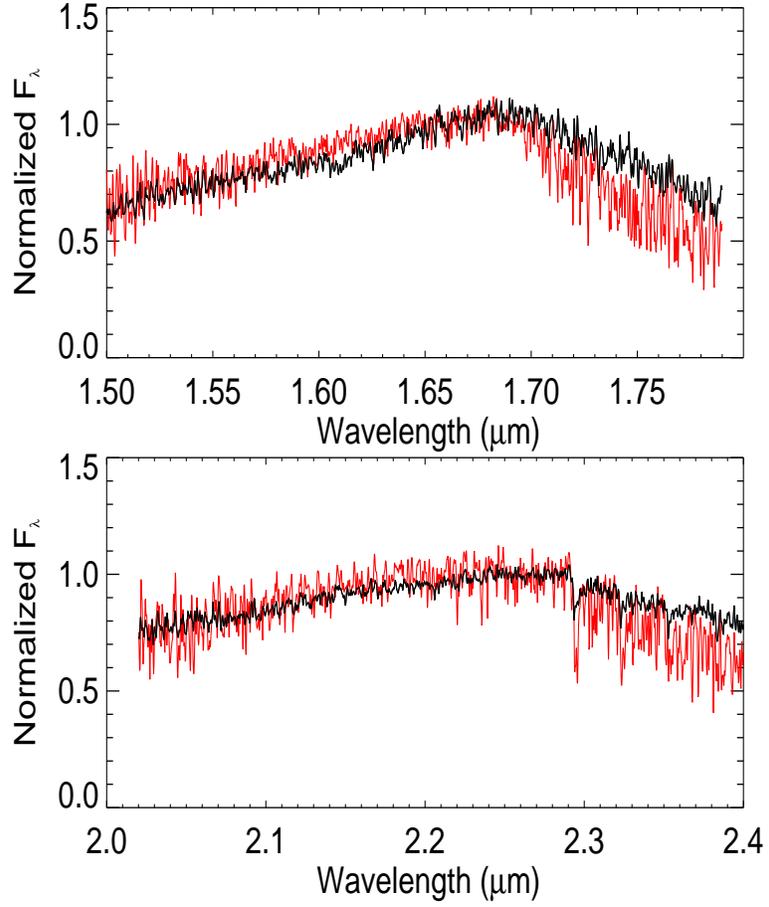}
     \caption{ Comparison of the GAIA synthetic spectra (red lines) at $T_{\rm{eff}}$ = 2400 K and $\log~ g = 4.0$ dex with GQ Lup b (black line) spectra in the $H$ and $K$ band at $R~\approx ~5000$. The spectra were normalized at 1.68 and 2.29 $\mu m$.   
    \label{fig:GaiaFitHK}}
  \end{center}
\end{figure}

\begin{figure}[!h]
  \begin{center}
    \includegraphics[width=12cm,height=12cm]{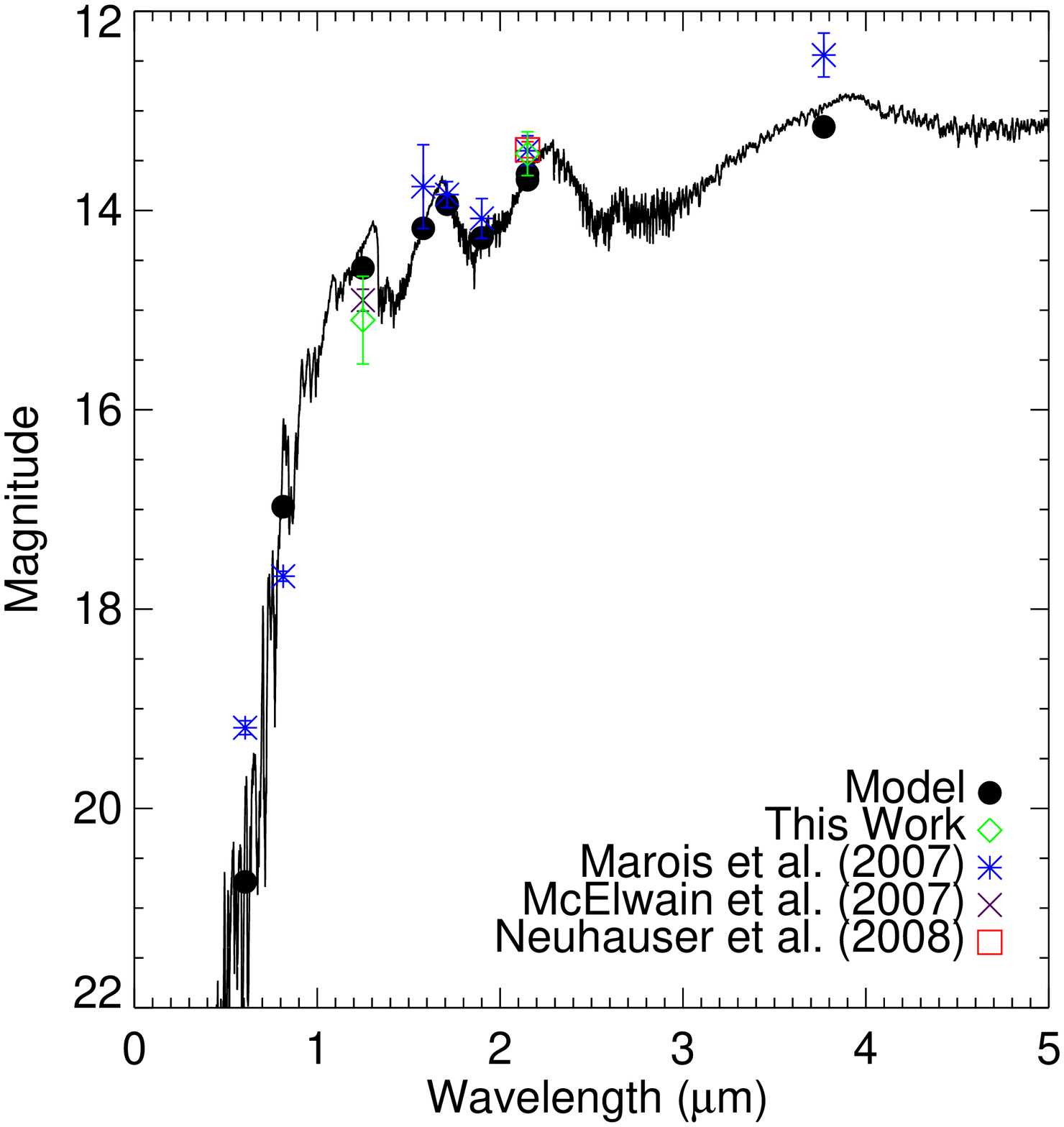}
     \caption{ Wide band GAIA synthetic spectra compared to photometric measurements taken in this work, in \citet{Marois07}, in \citet{McElwain07} and in \citet{Neuhauser08}. The model spectrum integrated flux in a given band is plotted as a black dot. As the measurements from this work and the $\rm CH_4$ and the $L^\prime$ values from \citet{Marois07} were made by differential photometry with the host star, the error bars were set to include the variability of GQ Lup A reported by \citet{Broeg07}. These error bars only reflect short term variations while long term fluctuations can also contribute to the discrepancy between predicted and measured values due to host star activity change.     
    \label{fig:compMag}}
  \end{center}
\end{figure}

\begin{figure}[!h]
  \begin{center}
    \includegraphics[width=12cm,height=12cm]{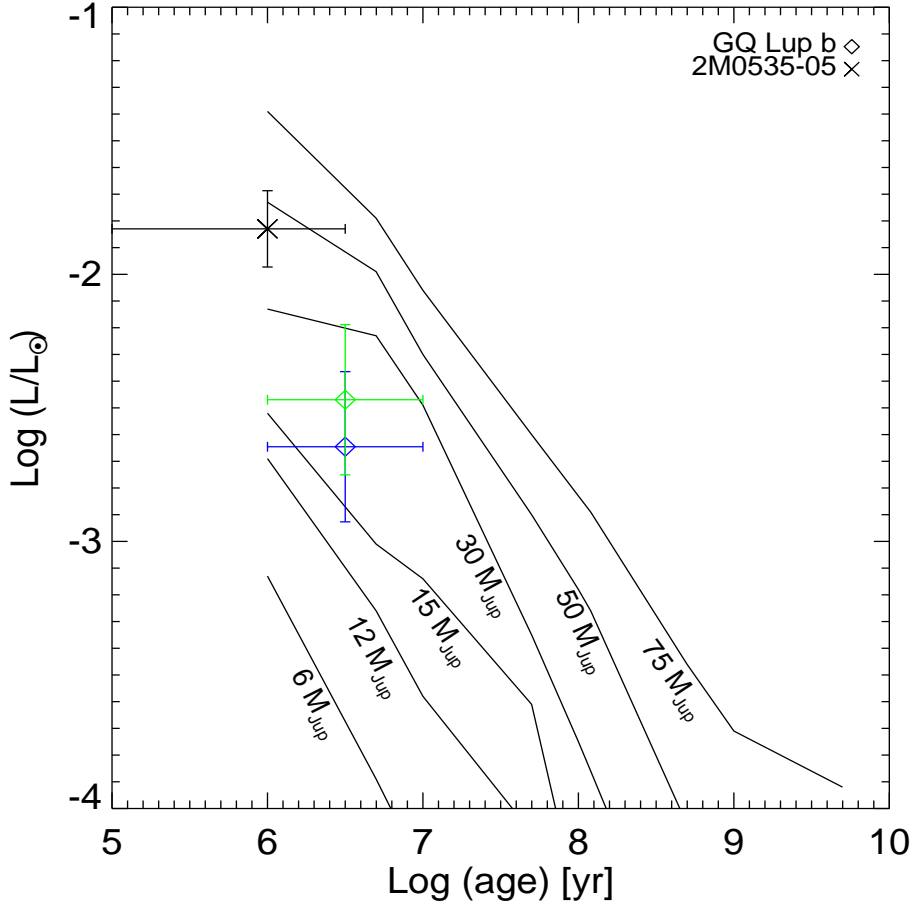}
     \caption{GQ Lup b position (diamond symbol) on the evolutionary tracks from \citet{Chabrier00} that begins at $10^6$ years. The luminosity uncertainty is introduced by the lack of precision in the system distance measurement. The lower mass object of the eclipsing binary brown dwarf 2M0535-05 that has a dynamically determined mass of $36 \pm 3$ $M_{\rm{Jup}}$ is also plotted as a reference point. The evolutionary model underestimate the luminosity of this object by a factor of $\sim 1.5$. Taking this systematic error into account leads to the blue error bars. 
    \label{fig:dustyLum}}
  \end{center}
\end{figure}

\end{document}